\documentclass[superscriptaddress,pra,twocolumn,floatfix]{revtex4}

\usepackage{amsmath,amsthm,amsfonts,amssymb,graphics,graphicx,epsfig,color,times,natbib,subfigure}

\newcommand{\braket}[2]{\mbox{$ \langle #1 | #2 \rangle $}}
\newcommand{\sandwich}[3]{\mbox{$ \langle #1 | #2 | #3 \rangle $}}
\newcommand{\ket}[1]{\mbox{$ | #1 \rangle $}}
\newcommand{\bra}[1]{\mbox{$ \langle #1 | $}}

\newcommand{\be}{\begin{equation}}
\newcommand{\ee}{\end{equation}}
\newcommand{\ba}{\begin{eqnarray}}
\newcommand{\ea}{\end{eqnarray}}

\newcommand{\si}{\sigma}
\newcommand{\demi}{\frac{1}{2}}
\newcommand{\compl}{\begin{picture}(8,8)\put(0,0){C}\put(3,0.3){\line(0,1){7}}\end{picture}}

\newcommand{\one}{\leavevmode\hbox{\small1\normalsize\kern-.33em1}}

\newcommand{\Tr}[1]{\mathrm{Tr} \ #1}

\begin{document}

\title{Evaluation of two different entanglement measures on a bound entangled state}

\author{Cyril Branciard}
\affiliation{Centre for Quantum Computer Technology, School of Mathematics and Physics, The University of Queensland, St Lucia, QLD 4072, Australia}
\author{Huangjun Zhu}
\affiliation{Centre for Quantum Technologies, National University of
Singapore, Singapore 117543, Singapore} \affiliation{NUS Graduate
School for Integrative Sciences and Engineering, Singapore 117597,
Singapore}
\author{Lin Chen}
\affiliation{Centre for Quantum Technologies, National University of
Singapore, Singapore 117543, Singapore}
\author{Valerio Scarani}
\affiliation{Centre for Quantum Technologies, National University of
Singapore, Singapore 117543, Singapore} \affiliation{Department of
Physics, National University of Singapore, Singapore 117542,
Singapore}
\date{\today}
\pacs{}

\begin{abstract}
We consider the mixed three-qubit bound entangled state defined as
the normalized projector on the subspace that is complementary to an
Unextendible Product Basis [C. H. Bennett {\it et. al.}, Phys. Rev.
Lett. {\bf 82}, 5385 (1999)]. Using the fact that no product state
lies in the support of that state, we compute its entanglement by
providing a basis of its subspace formed by ``minimally-entangled"
states. The approach is in principle applicable to any entanglement
measure; here we provide explicit values for both the geometric
measure of entanglement and a generalized concurrence.
\end{abstract}

\maketitle

\section{Introduction}


Entanglement measures quantify how much entanglement is contained in
a quantum state, which plays a fundamental role in quantum
information and computation tasks (for a recent review, see
\cite{hhh09}). For example, distillable entanglement quantifies how
many Bell states one can asymptotically obtain from target states
under local operations and classical communications (LOCC)
\cite{bds96}; in particular, there exist bound entangled states,
whose distillable entanglement vanishes under LOCC \cite{hhh98}.
Many entanglement measures have been proposed to characterize
multipartite states. These measures include the relative entropy of
entanglement \cite{vpr97}, the geometric measure of entanglement
\cite{wg03} and the generalized concurrence \cite{Mintert05}. Multipartite
entangled states are a useful resource for promising quantum information
tasks, such as one-way quantum computation \cite{rb01} and
multi-user quantum communications \cite{rl09}: it is therefore essential to be
able to characterize their entanglement. In practice, some
entanglement measures can be estimated from experimental data \cite{oh10}, or via
the efficient method of direct measurements, which can be turned into
a verification test in experiments \cite{vanenk09}.

However, entanglement measures are usually difficult to estimate,
especially for multipartite mixed states
\cite{wei04,hmm08,tam09,chen09,zch10,mgb10}.
In this paper we study the three-qubit mixed bound entangled state
defined as the normalized projector on the subspace complementary to
an Unextendible Product Basis (UPB)~\cite{UPB_PRL}. This is a
typical example of a multipartite mixed state whose entanglement can be
detected by entanglement witnesses built from local observables
\cite{aac10}. Recently, a scheme for studying local
distinguishability of three-qubit UPB states has also been proposed
\cite{dxy10}. Here, we use a unified strategy to compute two
measures, namely the geometric measure of entanglement and the
generalized concurrence.
We find in particular that the optimal
decompositions for both measures are different. Our strategy is quite
general, and could in principle be applied to all entanglement measures.

The paper is organized as follows. In Sec. II we
introduce the target state and its relation with an UPB, as well as the
strategy of computing the two entanglement measures in the following
sections. In Sec. III we analytically derive the geometric measure
of entanglement. In Sec. IV we give a numerical lower bound and an
analytical upper bound on the generalized concurrence. Finally we
conclude in Sec. V.

\section{The state and the strategy}

\subsection{Unextendible Product Bases (UPB) \\ and bound entanglement}
\label{ssupb}

Consider the following four three-qubit states: \be
    \begin{array}{rcccccccc}
        \ket{\varphi_0} &=& | & \!\!\! 0 & \!\!\! \rangle_A \ | & \!\!\! 0 & \!\!\! \rangle_B \ | & \!\!\! 0 & \!\!\! \rangle_C \\
            \ket{\varphi_1} &=& | & \!\!\! 1 & \!\!\! \rangle_A \ | & \!\!\! + & \!\!\! \rangle_B \ | & \!\!\! - & \!\!\! \rangle_C \\
            \ket{\varphi_2} &=& | & \!\!\! - & \!\!\! \rangle_A \ | & \!\!\! 1 & \!\!\! \rangle_B \ | & \!\!\! + & \!\!\! \rangle_C \\
            \ket{\varphi_3} &=& | & \!\!\! + & \!\!\! \rangle_A \ | & \!\!\! - & \!\!\! \rangle_B \ | & \!\!\! 1 & \!\!\! \rangle_C
    \end{array}\,.
    \label{eq_UPB}
\ee
They form an Unextendible Product Basis (UPB)~\cite{UPB_PRL}: no
product state can be found orthogonal to the four states
in~(\ref{eq_UPB}). In other words, if ${\cal P}$ is the subspace
generated by the four vectors (\ref{eq_UPB}), there is no product
state in the complementary space ${\cal Q}=\one-{\cal P}$.

The state defined as the uniform mixture on the space complementary to a UPB is always a bound entangled state \cite{UPB_PRL, UPBother}.
 In our case, this state reads \ba
    \rho_{{\cal Q}} & =& \frac{1}{4} \left( \one - \sum_{i=0}^{3}{\ket{\varphi_i}\bra{\varphi_i}} \right)\,
    \label{eq_def_rho}\\
    &=& \frac{1}{8} \Big[ \one - \frac{1}{2} \big( \one\otimes\si_+\otimes\si_- + (\mathrm{cyclic})\big) \nonumber\\ && - \frac{1}{2\sqrt{2}} (\si_+\otimes\si_+\otimes\si_+ + \si_-\otimes\si_-\otimes\si_-) \Big] \label{explrho}
\ea with $\si_\pm = (\si_z \pm \si_x)/\sqrt{2}$ \cite{note1}. It is convenient to review rapidly its remarkable properties.

By definition, $\rho_{{\cal Q}}$ is entangled: there is no product state in its support, so \textit{a fortiori} it will be impossible to decompose it on product states. It can also be easily verified \cite{UPB_PRL} that one can complete the basis (\ref{eq_UPB}) with four vectors such that the first two qubits are entangled and the third one is separable; in other words, $\rho_{{\cal Q}}$ can be decomposed in the form
\ba
\rho_{{\cal Q}}=\frac{1}{4}\sum_{i=0}^3 \ket{\Psi_i}_{AB}\bra{\Psi_i}\otimes\ket{\Psi_i'}_{C}\bra{\Psi_i'}\,.\label{decomp}
\ea
This means that the state is not three-partite entangled.
Moreover, from this decomposition it is obvious that the reduced states $\rho_{AC}$ and $\rho_{BC}$ are separable;
 even more: no measurement of $B$ can prepare an entangled state between $A$ and $C$,
 and no measurement of $A$ can prepare an entangled state between $B$ and $C$. At first sight, one might hope that $\rho_{AB}$ is entangled,
 or at least that a measurement of $C$ could prepare an entangled state between $A$ and $B$.
 However, this is not the case. Indeed, by construction of the basis (\ref{eq_UPB}), $\rho_{{\cal Q}}$ is invariant
 under cyclic permutations $A\rightarrow B\rightarrow C\rightarrow A$. Therefore, we can rewrite (\ref{decomp})
 with states $\ket{\Psi_i}_{BC}\ket{\Psi_i'}_{A}$ and repeat the reasoning above. In conclusion, the entanglement
 that has to be invested to create $\rho_{{\cal Q}}$ is nowhere to be found back, whichever partition and LOCC strategy is envisaged.

The state $\rho_{{\cal Q}}$ is thus a paradigmatic example of bound
entanglement. Since it is not symmetric under all permutations, but
only the cyclic ones, it does not fall in the family of states for
which general studies of entanglement measures have been made~\cite{hub09,markham10,zch10,mgb10}.
Here we shall show how one can compute the value of entanglement for
such a state. The main idea is presented in the next paragraph.

\subsection{Quantifying entanglement}

Entanglement measures are normally defined on pure states. For bipartite pure states, there is only one measure, namely the entropy of the reduced density matrix. For multipartite states, the situation is much less well understood and there are several candidates for entanglement measures. Still, they are usually computable on pure states.

The definition of an entanglement measure $E$ on pure states can be extended to any mixed state $\rho$ as follows:
\be
    E(\rho) = \min_{\{p_i,\ket{\psi_i}\}} \sum_i p_i
    E(\ket{\psi_i}),
    \label{eq_Esigma}
\ee
where the minimum is to be taken among all possible pure-state decompositions of $\rho$ in the form $\rho = \sum_i p_i \ket{\psi_i}\bra{\psi_i}$. There is however no general recipe to compute this minimum.

An obvious lower bound is given by
\be
    E(\rho) \geq \min_{\ket{\psi} \in \mathrm{supp}(\rho)} E(\ket{\psi}), \label{lower}
\ee
where $\mathrm{supp}(\rho)$ is the support of $\rho$. For many mixed states, this lower bound is trivial, because there will be a product state in the support of $\rho$ and therefore the r.h.s. is simply 0. However, \textit{for $\rho_{{\cal Q}}$ the r.h.s. of (\ref{lower}) is not zero}, because we know that there is no product state in its support.

In addition, for the two measures of entanglement considered below,
we shall show that one can find a complete orthonormal
\textit{basis} $\{\ket{\psi_0},\dots,\ket{\psi_3}\}$ of ${\cal Q}$
formed by ``minimally-entangled" states, ie. such that \be
E(\ket{\psi_i})=\min_{\ket{\psi} \in {\cal Q}} E(\ket{\psi})\ee for
all $i=0,\dots,3$. As $\rho_{\cal Q} =\frac{1}{4}
\sum\ket{\psi_i}\bra{\psi_i}$ for any orthonormal basis
$\{\ket{\psi_i}\}$ of ${\cal Q}$, this implies that, for these two
measures at least, \ba E(\rho_{\cal Q}) &=& \min_{\ket{\psi} \in
{\cal Q}} E(\ket{\psi})\,.\label{lowerQ} \ea This is the simple but
crucial insight that will allow us to compute the entanglement of
$\rho_{{\cal Q}}$. A similar insight was used for the Smolin state,
a permutation invariant four-qubit bound entangled state
\cite{wei04}. In the next two sections, we compute the r.h.s. of
(\ref{lowerQ}) and exhibit a full basis that reaches this value, for
two measures of multipartite entanglement, the geometric measure of
entanglement \cite{wg03} in Sec.~\ref{secgeom} and a generalized
concurrence \cite{Mintert05} in Sec.~\ref{secmint}.

\section{Geometric entanglement of $\rho_{\cal Q}$}
\label{secgeom}

We start by considering the geometric measure of entanglement \cite{wg03}. For an $N$-partite pure state $\ket{\psi}$, this measure is defined as
\be
    E_G(\ket{\psi}) = 1-\max_{\ket{\phi} \in
    \Pi}|\braket{\psi}{\phi}|^2,
    \label{def_EG}
\ee
where $\Pi$ is the set of all $N$-partite pure product states $\ket{\phi} = \ket{\phi_1}\otimes\dots\otimes\ket{\phi_N}$.

Following the strategy defined above, we are going to compute
\ba
    E_G(\rho_{\cal Q}) &=& 1-\frac{3\sqrt{6}}{8} \simeq 0.08144.
  \label{eq_EG_rho}
\ea For comparison, in the case of three qubits, the largest value
of geometric measure of entanglement is achieved for the $W$ state
and is $E_G(\ket{W})=\frac{5}{9}$ \cite{tam09,chen09}.

\subsection{Calculating $\min_{\ket{\psi} \in {\cal Q}} E_G(\ket{\psi})$}

From the definition (\ref{def_EG}) of $E_G$, we have
\ba
\min_{\ket{\psi} \in {\cal Q}} E_G(\ket{\psi}) &=& \min_{\ket{\psi} \in {\cal Q}} (1-\max_{\ket{\phi} \in \Pi}|\braket{\psi}{\phi}|^2) \nonumber \\
&=& 1 - \max_{\ket{\phi} \in \Pi} \max_{\ket{\psi} \in {\cal Q}} |\braket{\psi}{\phi}|^2.
\ea

Now, for a given $\ket{\phi} \in \Pi$, the closest state to $\ket{\phi}$ in the subspace ${\cal Q}$ is simply the projection of $\ket{\phi}$ onto ${\cal Q}$. Denoting by $\tilde {\cal Q}$ the projector onto the subspace ${\cal Q}$, we get
\be
\max_{\ket{\psi} \in {\cal Q}} |\braket{\psi}{\phi}|^2 = \sandwich{\phi}{\tilde {\cal Q}}{\phi},
\ee
so that
\ba
\min_{\ket{\psi} \in {\cal Q}} E_G(\ket{\psi}) &=& 1 - \max_{\ket{\phi} \in \Pi} \sandwich{\phi}{\tilde {\cal Q}}{\phi} \nonumber \\
&=& \min_{\ket{\phi} \in \Pi} \sandwich{\phi}{\tilde {\cal
P}}{\phi}, \ea where $\tilde {\cal P}$ denotes the projector onto
the subspace ${\cal P}$. We show in Appendix~\ref{App_EG} how to
calculate $\min_{\ket{\phi} \in \Pi} \sandwich{\phi}{\tilde {\cal
P}}{\phi}$ analytically. We obtain \be \min_{\ket{\psi} \in {\cal
Q}} E_G(\ket{\psi}) = 1-\frac{3\sqrt{6}}{8}. \label{eq_EG_LB} \ee

\subsection{A whole basis reaching $\min_{\ket{\psi} \in {\cal Q}} E_G(\ket{\psi})$}

As also shown in Appendix~\ref{App_EG}, the minimum $\min_{\ket{\phi} \in \Pi} \sandwich{\phi}{\tilde {\cal P}}{\phi}$ can be attained by the four states $\ket{\phi_i} = \ket{a_i}\ket{b_i}\ket{c_i}$ ($i=0,\dots,3$) with
\ba
\ket{a_i} &=& \cos \frac{\alpha_i}{2} \ket{0} + \sin \frac{\alpha_i}{2} \ket{1} \\
\ket{b_i} &=& \cos \frac{\beta_i}{2} \ket{0} + \sin \frac{\beta_i}{2} \ket{1} \\
\ket{c_i} &=& \cos \frac{\gamma_i}{2} \ket{0} + \sin \frac{\gamma_i}{2} \ket{1}
\ea
and for the following values of $\alpha_i, \beta_i, \gamma_i$~\cite{footnote_symmetry}:
\ba
\begin{array}{llll}
\ket{\phi_0}: \ & \alpha_0 = \theta_0, & \beta_0 = \theta_0, & \gamma_0 = \theta_0, \\
\ket{\phi_1}: & \alpha_1 = \pi+\theta_0, & \beta_1 = \frac{\pi}{2}-\theta_0, & \gamma_1 = \frac{3\pi}{2}-\theta_0, \\
\ket{\phi_2}: & \alpha_2 = \frac{3\pi}{2}-\theta_0, & \beta_2 =\pi+\theta_0, & \gamma_2 = \frac{\pi}{2}-\theta_0, \\
\ket{\phi_3}: & \alpha_3 = \frac{\pi}{2}-\theta_0, & \beta_3 =
\frac{3\pi}{2}-\theta_0, & \gamma_3 =\pi+\theta_0,
\end{array}
\ea where $\theta_0 = \arccos (-\frac{\sqrt{6}-2}{2})$.

Let $\ket{\psi_i}$ ($i=0,\dots,3$) be the normalized states corresponding to the projection of $\ket{\phi_i}$ onto ${\cal Q}$:
\be
\ket{\psi_i} = \frac{1}{\sqrt{\frac{3\sqrt{6}}{8}}} \ \tilde {\cal Q} \ket{\phi_i}.
\ee
On the one hand, it can be checked that these four states form an orthonormal basis of ${\cal Q}$; they thus provide a decomposition $\rho_{\cal Q} = \frac{1}{4}\sum_{i=0}^3 \ket{\psi_i}\bra{\psi_i}$. On the other hand, by construction
\ba
E_G(\ket{\psi_i}) &=& 1-\max_{\ket{\phi} \in \Pi}|\braket{\psi_i}{\phi}|^2 = 1-|\braket{\psi_i}{\phi_i}|^2 \nonumber \\
&=& 1-\frac{3\sqrt{6}}{8}. \ea This concludes the proof of
(\ref{eq_EG_rho}).

As a remark: 
the four states $\ket{\psi_i}$ turn out to be \textit{three-partite}
entangled. As we mentioned in section \ref{ssupb}, there exist bases
of ${\cal Q}$ made of bipartite entangled states. It is interesting
to note that such bases are not those that minimize the geometric
measure of entanglement.

\section{Generalized concurrence of $\rho_{\cal Q}$}
\label{secmint}

The second measure of entanglement that we consider is a generalization of the concurrence defined as \cite{Mintert05}
\be
    E_C(\ket{\psi}) = 2^{1-N/2} \sqrt{2^N-2-\sum_j{\Tr{\rho_j^2}}},
    \label{def_EC}
\ee
where the multi-index $j$ runs over all $(2^N-2)$ subsets of the $N$ subsystems and $\rho_j$ is the reduced density matrix of the corresponding subset.

Following the strategy defined above, we are going to prove that
\ba
    E_C(\rho_{\cal Q}) &=& \frac{\sqrt{897}}{52} \simeq 0.57596\,.
  \label{eq_EC_rho}
\ea In fact, analytically we will prove only $E_C(\rho)\leq\frac{\sqrt{897}}{52}$, but we have strong numerical evidence that this is indeed the exact value of $E_C(\rho)$.

In comparison, in the case of three qubits, the largest value of this measure of entanglement is reached for the $GHZ$ state (for which all the reduced states $\rho_j$ in (\ref{def_EC}) are maximally mixed) and is $E_C(\ket{GHZ}) = \sqrt{3/2} \simeq 1.2247$.

\subsection{Calculating $\min_{\ket{\psi} \in {\cal Q}} E_C(\ket{\psi})$}

Consider first states that lie in the symmetric subspace of
${(\compl^2)}^{\otimes 3}$, denoted ${\cal S}$. Among these states,
we can find analytically the one that minimizes $E_C(\ket{\psi})$.
The analytical calculations are detailed in Appendix~\ref{App_EC},
and the final result is \be \min_{\ket{\psi} \in {\cal Q} \cap {\cal
S}} E_C(\ket{\psi}) = \frac{\sqrt{897}}{52}. \ee We have not been
able to prove analytically that this value defines the minimum of
$E_C(\ket{\psi})$ over the whole subspace ${\cal Q}$; however, a
brute force numerical minimization reaches exactly the same value.
Therefore, up to the conjecture that there exists a symmetric state
$\ket{\psi_0'}$ that reaches the minimum, backed by numerical
evidence, we can assert that \be \min_{\ket{\psi} \in {\cal Q}}
E_C(\ket{\psi}) = \frac{\sqrt{897}}{52}\,. \label{eq_EC_LB} \ee

\subsection{A whole basis reaching $\min_{\ket{\psi} \in {\cal Q}} E_G(\ket{\psi})$}

We also exhibit in Appendix~\ref{App_EC} three other states $\ket{\psi_1'}, \ket{\psi_2'}$ and $\ket{\psi_3'}$,
 that form an orthonormal basis of ${\cal Q}$ together with $\ket{\psi_0'}$, and that are such that
\ba E_C(\ket{\psi_1'}) &=& E_C(\ket{\psi_2'}) = E_C(\ket{\psi_3'}) =
\frac{\sqrt{897}}{52}\nonumber\\
& =& E_C(\ket{\psi_0'})\,. \ea Up to the conjecture mentioned above,
this concludes the proof of (\ref{eq_EC_LB}).

Note that, as it was the case for geometric measure of entanglement,
all these four states are three-partite entangled.

\section{Conclusion}

We found a way to estimate the entanglement of the state $\rho_{\cal
Q}$. It is the first time the entanglement of this state is
quantified in terms of geometric measure of entanglement and
generalized concurrence \cite{ys06}; and it is found to be strictly positive, while the state is
bound-entangled and not fully three-partite entangled.

The remarkable property of $\rho_{\cal Q}$ that allowed us to estimate its entanglement is the possibility to decompose it into a mixture of ``minimally-entangled'' states. This was at least possible for both the geometric measure of entanglement and the generalized concurrence; we don't know whether this necessarily holds for all entanglement measures. Nevertheless, our results also illustrate the fact that the two measures of entanglement that we considered are quite different: the optimal
decomposition of $\rho_{\cal Q}$ as a mixture of pure states is not
the same for these two measures of entanglement.


Our technique can be applied to other states, whenever such a decomposition into ``minimally-entangled'' states is possible. As further examples, we give in Appendix~\ref{App_GenShifts} numerical results for the bound entangled state constructed out of the generalized three-qubit ``GenShifts" UPB~\cite{UPB_PRL, UPBother}. It would also be interesting to apply our approach to states with more parties, and to different entanglement measures.


\section{Acknowledgments}

We acknowledge discussions with Robert H\"ubener and Andreas Winter.
This work was supported by the Australian Research Council Centre of Excellence for Quantum Computer Technology and by the National Research Foundation and the
Ministry of Education, Singapore.

\appendix


\section{Calculations for $E_G(\rho_{\cal Q})$: \newline
A{\lowercase{nalytical calculation of}} $\min_{\ket{\phi} \in \Pi} \sandwich{\phi}{\tilde {\cal P}}{\phi} \qquad $}

\label{App_EG}


Any state $\ket{\phi} \in \Pi$ can be written as $\ket{\phi} = \ket{a}\ket{b}\ket{c}$, with
\ba
 \ket{a} &=& \cos \frac{\alpha}{2} \ket{0} + \sin \frac{\alpha}{2} e^{i \varphi_a} \ket{1} \\
\ket{b} &=& \cos \frac{\beta}{2} \ket{0} + \sin \frac{\beta}{2} e^{i \varphi_b} \ket{1} \\
 \ket{c} &=& \cos \frac{\gamma}{2} \ket{0} + \sin \frac{\gamma}{2} e^{i \varphi_c} \ket{1}
\ea
and where $\alpha, \beta, \gamma \in [0,2\pi]$; $\varphi_a, \varphi_b, \varphi_c \in [0,\pi]$. With these notations, we find
\ba
\sandwich{\phi}{\tilde {\cal P}}{\phi} && \!\!\!\! = \ \ \frac{1}{8}(1+\cos\alpha)(1+\cos\beta)(1+\cos\gamma) \nonumber \\
 && + \frac{1}{8}(1-\cos\alpha)(1+\sin\beta\cos\varphi_b)(1-\sin\gamma\cos\varphi_c) \nonumber \\
 && + \frac{1}{8}(1-\sin\alpha\cos\varphi_a)(1-\cos\beta)(1+\sin\gamma\cos\varphi_c) \nonumber \\
 && + \frac{1}{8}(1+\sin\alpha\cos\varphi_a)(1-\sin\beta\cos\varphi_b)(1-\cos\gamma) . \nonumber \\
\ea
The above expression being linear in $\cos{\varphi_a}$, its minimum can be attained for either $\varphi_a = 0$ or $\varphi_a = \pi$. As the expression is also invariant under the transformation $(\alpha \leftrightarrow 2\pi-\alpha, \varphi_a \leftrightarrow \pi-\varphi_a)$, then its minimum can be attained for $\varphi_a = 0$.

Similar arguments can be applied to $\varphi_b$ and $\varphi_c$,
which allows us to write \ba \min_{\ket{\phi} \in \Pi}
\sandwich{\phi}{\tilde {\cal P}}{\phi} = \min_{\alpha, \beta, \gamma
\in [0,2\pi]} F(\alpha, \beta, \gamma), \ea where \ba
F(\alpha, \beta, \gamma) &=& \frac{1}{8}(1+\cos\alpha)(1+\cos\beta)(1+\cos\gamma) \nonumber \\
 && + \frac{1}{8}(1-\cos\alpha)(1+\sin\beta)(1-\sin\gamma) \nonumber \\
 && + \frac{1}{8}(1-\sin\alpha)(1-\cos\beta)(1+\sin\gamma) \nonumber \\
 && + \frac{1}{8}(1+\sin\alpha)(1-\sin\beta)(1-\cos\gamma) \nonumber \\
 &=& 1 - \frac{1}{16} \det M
\ea
with
\ba
M = \left(
 \begin{array}{ccc}
 \cos\alpha - \sin\alpha & \cos\alpha + \sin\alpha & -2 \\
 \cos\beta + \sin\beta & -2 & \cos\beta - \sin\beta \\
 -2 & \cos\gamma - \sin\gamma & \cos\gamma + \sin\gamma
 \end{array}\right).
 \nonumber
\ea
Now, the Hadamard inequality applied to the row vectors of $M$ gives
\be
|\det M| \leq 6\sqrt{6},
\ee
from which we conclude that
\ba
F(\alpha, \beta, \gamma) \geq 1-\frac{3\sqrt{6}}{8}.
\ea

The equality is obtained if the three row vectors of $M$ are
mutually orthogonal. The following four sets of values for $\alpha,
\beta, \gamma$, with $\cos\theta_0 = -\frac{\sqrt{6}-2}{2}$, all
satisfy this condition: \ba
\begin{array}{llll}
\ket{\phi_0}: \ & \alpha_0 = \theta_0, & \beta_0 = \theta_0, & \gamma_0 = \theta_0, \\
\ket{\phi_1}: & \alpha_1 = \pi+\theta_0, & \beta_1 = \frac{\pi}{2}-\theta_0, & \gamma_1 = \frac{3\pi}{2}-\theta_0, \\
\ket{\phi_2}: & \alpha_2 = \frac{3\pi}{2}-\theta_0, & \beta_2 =\pi+\theta_0, & \gamma_2 = \frac{\pi}{2}-\theta_0, \\
\ket{\phi_3}: & \alpha_3 = \frac{\pi}{2}-\theta_0, & \beta_3 =
\frac{3\pi}{2}-\theta_0, & \gamma_3 =\pi+\theta_0,
\end{array}
\ea Interestingly, the four states $\ket{\phi_i}$ defined by the
corresponding values of  $\alpha_i, \beta_i, \gamma_i$ (and with
$\varphi_a=\varphi_b=\varphi_c=0$) also form a UPB, of the
``GenShifts" type~\cite{UPB_PRL, UPBother}.

These four states all attain the previous lower bound for $F(\alpha, \beta, \gamma)$, so that for $i=0,\dots,3$,
\ba
\sandwich{\phi_i}{\tilde {\cal P}}{\phi_i} = \min_{\ket{\phi} \in \Pi} \sandwich{\phi}{\tilde {\cal P}}{\phi} = 1 - \frac{3\sqrt{6}}{8}.
\ea

\section{Calculations for $E_C(\rho_{\cal Q})$}
\label{App_EC}

\subsection{Calculation of $\min_{\ket{\psi} \in {\cal Q} \cap {\cal S}} E_C(\ket{\psi})$}

For convenience, let us start by defining a basis for ${\cal Q}$. A possible choice is the following:
\be
    \begin{array}{rclcccccccc}
        \ket{q_0} &=& \frac{1}{\sqrt{2}}(| &\!\!\!+&\!\!\!+&\!\!\!+&\!\!\!\rangle-| &\!\!\!-&\!\!\!-&\!\!\!-&\!\!\!\rangle) \\
        \ket{q_1} &=& \frac{1}{\sqrt{2}}(| &\!\!\!+&\!\!\!1&\!\!\!0&\!\!\!\rangle-| &\!\!\!-&\!\!\!0&\!\!\!1&\!\!\!\rangle) \\
        \ket{q_2} &=& \frac{1}{\sqrt{2}}(| &\!\!\!0&\!\!\!+&\!\!\!1&\!\!\!\rangle-| &\!\!\!1&\!\!\!-&\!\!\!0&\!\!\!\rangle) \\
        \ket{q_3} &=& \frac{1}{\sqrt{2}}(| &\!\!\!1&\!\!\!0&\!\!\!+&\!\!\!\rangle-| &\!\!\!0&\!\!\!1&\!\!\!-&\!\!\!\rangle)
    \end{array}\,.
    \label{eq_Q_basis}
\ee

A generic symmetric state in ${\cal Q}$ can then be written as
\ba
    \ket{\psi} = \cos\theta \ket{q_0} + \sin\theta e^{i \gamma} \frac{\ket{q_1} + \ket{q_2} + \ket{q_3}}{\sqrt{3}}
\ea
with $\theta,\gamma \in [0,\pi]$.

The sum in the definition (\ref{def_EC}) of $E_C(\ket{\psi})$ can be explicitly calculated and is found to be
\ba
        \begin{array}{r}
        \sum_j{\Tr{\rho_j^2}} = \frac{1}{3}(10 + 25 \cos^2\theta - 26 \cos^4\theta) \quad \\
                - 4 \cos^2\theta\sin^2\theta (1-\cos2 \gamma)\,.
    \end{array}
\ea The maximum of this two-variable function is obtained for
$\cos^2\theta=\frac{25}{52}, \cos2\gamma=1$. We get \be
\min_{\ket{\psi} \in {\cal Q} \cap {\cal S}} E_C(\ket{\psi}) =
\frac{\sqrt{897}}{52}, \ee and the minimum can be reached by two
different symmetric states, one of these being \be \ket{\psi_0'} =
\frac{1}{2\sqrt{13}}(5\ket{q_0}+3\ket{q_1}+3\ket{q_2}+3\ket{q_3}).
\ee

\subsection{Three other orthogonal states with the same value of $E_C$}

One can easily check that the following three states
\ba
    \begin{array}{rcl}
      \ket{\psi_1'} &=& \frac{1}{2\sqrt{13}}(3\ket{q_0}-5\ket{q_1}+3\ket{q_2}-3\ket{q_3}) \\
      \ket{\psi_2'} &=& \frac{1}{2\sqrt{13}}(3\ket{q_0}-3\ket{q_1}-5\ket{q_2}+3\ket{q_3}) \\
      \ket{\psi_3'} &=& \frac{1}{2\sqrt{13}}(3\ket{q_0}+3\ket{q_1}-3\ket{q_2}-5\ket{q_3})
    \end{array}
\ea
all have the same value of $E_C$ as $\ket{\psi_0'}$, and that, together with $\ket{\psi_0'}$, they form an orthonormal basis of ${\cal Q}$; they thus provide a decomposition of $\rho_{\cal Q}$.

\bigskip

\section{Entanglement of the bound-entangled state constructed out of the three-qubit ``GenShifts" UPB}

\label{App_GenShifts}

In this Appendix we apply our approach to the three-qubit bound-entangled state $\rho_{\cal Q}(\phi)$ constructed out of the generalized ``GenShifts" UPB~\cite{UPB_PRL, UPBother}, and show numerical results for the values of $E_G(\rho_{\cal Q}(\phi))$ and $E_C(\rho_{\cal Q}(\phi))$.

For a single-qubit state $\ket{\phi}$ and its orthogonal state $\ket{\phi^\perp}$, the GenShifts UPB consists of the four states
\be
    \begin{array}{rcccc}
        \ket{\varphi_0} &=& | & \!\!\! 0, \ 0, \ 0 & \!\!\! \rangle \\
            \ket{\varphi_1} &=& | & \!\!\! 1,\phi,\phi^\perp & \!\!\! \rangle \\
            \ket{\varphi_2} &=& | & \!\!\! \phi^\perp,1,\phi & \!\!\! \rangle \\
            \ket{\varphi_3} &=& | & \!\!\! \phi,\phi^\perp,1 & \!\!\! \rangle
    \end{array}\,.
    \label{eq_UPB_GenShifts}
\ee
$\rho_{{\cal Q}}(\phi)$ is then defined as the uniform mixture on the complementary subspace ${\cal Q}(\phi)$:
\ba
    \rho_{{\cal Q}}(\phi) & =& \frac{1}{4} \left( \one - \sum_{i=0}^{3}{\ket{\varphi_i}\bra{\varphi_i}} \right)\,.
\ea
Note that the state $\rho_{\cal Q}$ studied in the main text is a particular case of $\rho_{\cal Q}(\phi)$, corresponding to $\ket{\phi}=\ket{+}$.

We checked numerically that the critical properties of $\rho_{\cal Q}$ that we used to estimate its entanglement are still satisfied by $\rho_{\cal Q}(\phi)$: for all choices of $\ket{\phi}$, one can again find two (different) orthonormal bases $\{\ket{\psi_i}\}$ and $\{\ket{\psi_i'}\}$ of ${\cal Q}(\phi)$ formed by ``minimally entangled" states,
ie. such that
\ba
 E_G(\ket{\psi_i}) &=& \min_{\ket{\psi} \in {\cal Q}(\phi)} E_G(\ket{\psi}), \\
 E_C(\ket{\psi_i'})&=& \min_{\ket{\psi} \in {\cal Q}(\phi)} E_C(\ket{\psi})
\ea
for all $i=0,\dots,3$. We then conclude that
\ba
 E_G(\rho_{\cal Q}(\phi)) &=& \min_{\ket{\psi} \in {\cal Q}(\phi)} E_G(\ket{\psi}), \\
 E_C(\rho_{\cal Q}(\phi)) &=& \min_{\ket{\psi} \in {\cal Q}(\phi)}
E_C(\ket{\psi}). \ea

\begin{figure}%
\includegraphics[width=8cm]{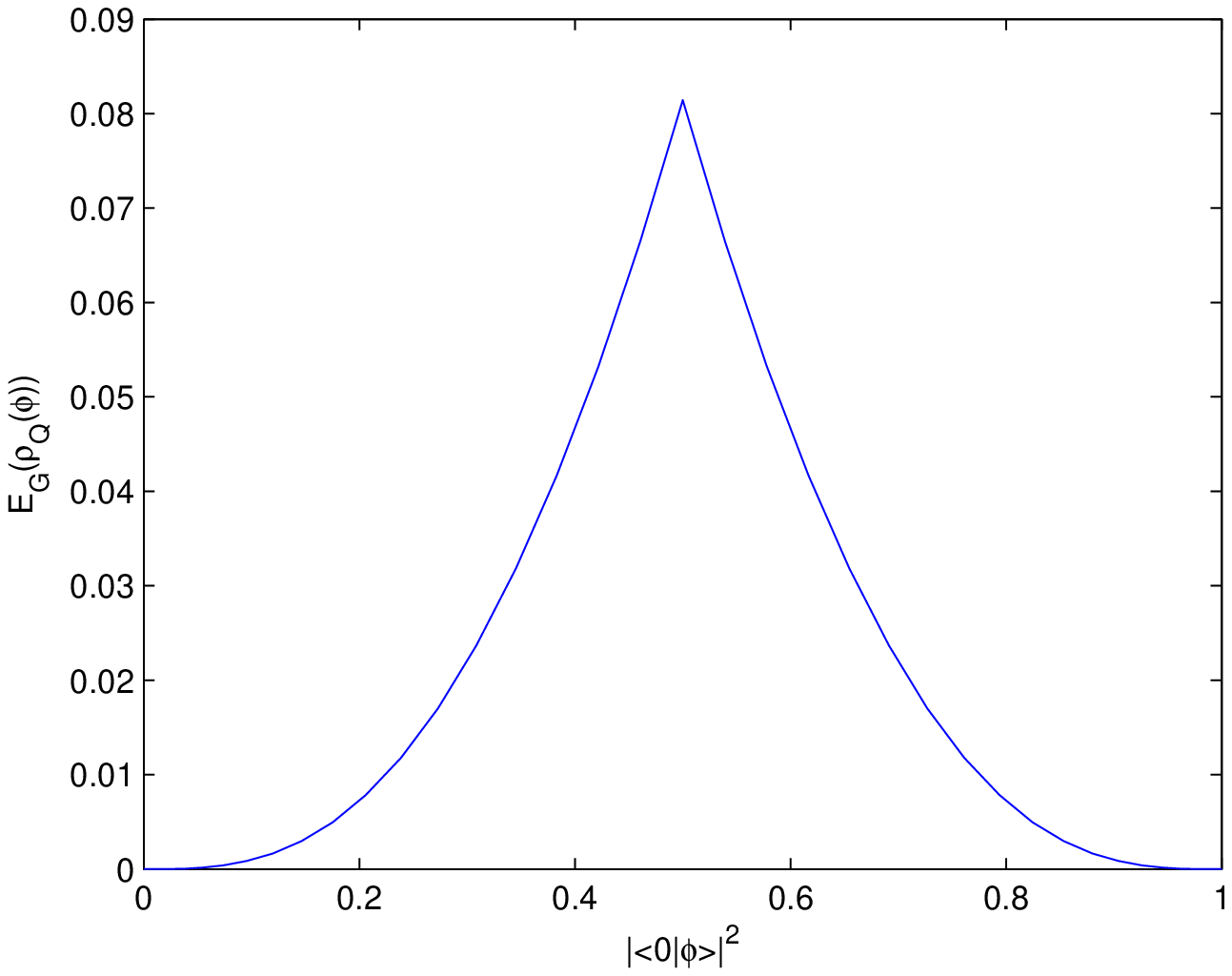}\\%
\includegraphics[width=8cm]{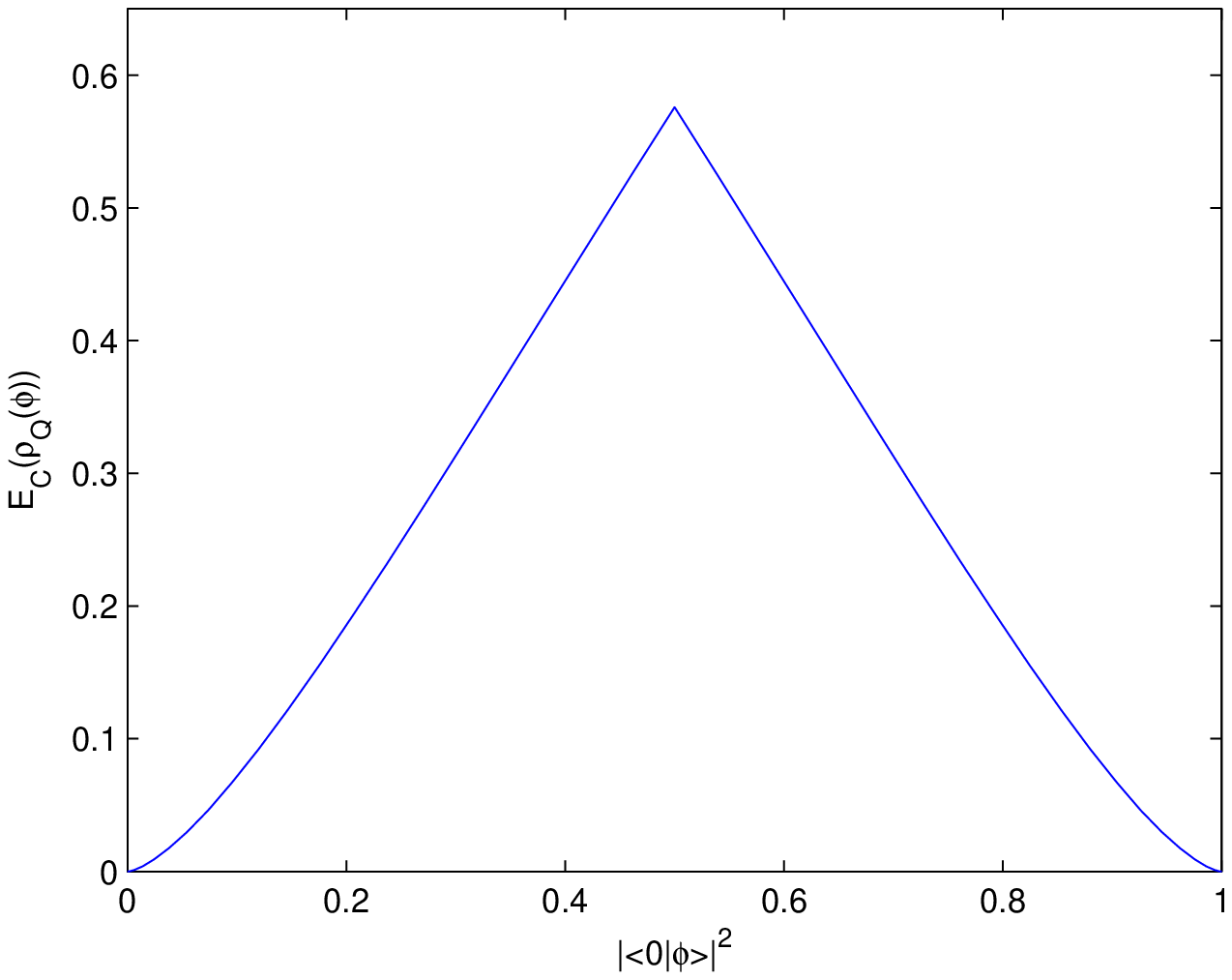}%
\caption{Numerical calculations of the values of $E_G(\rho_{\cal Q}(\phi))$ (top) and $E_C(\rho_{\cal Q}(\phi))$ (bottom).}%
\label{fig_GenShifts}%
\end{figure}

Figure~\ref{fig_GenShifts} displays the numerical results we obtained for the entanglement of $\rho_{\cal Q}(\phi)$ measured by $E_G$ and $E_C$, as a function of the overlap $|\braket{0}{\phi}|^2$. Not surprisingly, the maxima of $E_G(\rho_{\cal Q}(\phi))$ and $E_C(\rho_{\cal Q}(\phi))$ are found when $|\braket{0}{\phi}|^2=|\braket{1}{\phi}|^2=\demi$, which is in particular the case for $\rho_{\cal Q}=\rho_{\cal Q}(+)$.

\begin{thebibliography}{99}

\bibitem{hhh09} R. Horodecki, P. Horodecki, M. Horodecki, and K. Horodecki, Rev. Mod. Phys. {\bf
81}, 865 (2009).

\bibitem{bds96} C. H. Bennett, D. P. DiVincenzo, J. A.
Smolin, W. K. Wootters, Phys. Rev. A {\bf54}, 3824 (1996).

\bibitem{hhh98} M. Horodecki, P. Horodecki, and R. Horodecki, Phys. Rev. Lett.
{\bf80}, 5239 (1998).

\bibitem{vpr97} V. Vedral, M. B. Plenio, M. A. Rippin, and P. L.
Knight, Phys. Rev. Lett. {\bf78}, 2275 (1997).

\bibitem{wg03} T.-C. Wei, and P. M. Goldbart, Phys. Rev. A {\bf68}, 042307 (2003).

\bibitem{Mintert05} F. Mintert, M. Ku\'s, and A. Buchleitner, Phys. Rev. Lett. {\bf 95}, 260502 (2005).

\bibitem{rb01} R. Raussendorf and H. J. Briegel, Phys. Rev. Lett. {\bf 86}, 5188 (2001).

\bibitem{rl09} T. C. Ralph and P. K. Lam, Nature Photonics {\bf 3}, 671 (2009).

\bibitem{oh10} A. Osterloh and P. Hyllus, Phys. Rev. A \textbf{81}, 022307
(2010).

\bibitem{vanenk09} S. J. van Enk, Phys. Rev. Lett. {\bf 102}, 190503 (2009).

\bibitem{wei04} T.-C. Wei, J. B. Altepeter, P. M. Goldbart, and W. J. Munro, Phys. Rev. A {\bf70}, 022322 (2004).

\bibitem{hmm08} M. Hayashi, D. Markham, M. Murao, M. Owari, S. Virmani, Phys. Rev. A \textbf{77},
012104 (2008).

\bibitem{tam09} S. Tamaryan,  T.-C. Wei, and D. Park,  Phys. Rev. A {\bf  80}, 052315 (2009).

\bibitem{chen09} L. Chen, A. Xu and H. Zhu, arXiv:0911.1493v1 (2009).

\bibitem{zch10} H. Zhu, L. Chen, and M. Hayashi, arXiv:
1002.2511 (2010).

\bibitem{mgb10} J. Martin, O. Giraud, P. A. Braun, D. Braun, and T. Bastin, arXiv:
1003.0593 (2010).

\bibitem{UPB_PRL} C. H. Bennett, D. P. DiVincenzo, T. Mor, P. W. Shor, J. A. Smolin, and B. M. Terhal, Phys. Rev. Lett. {\bf 82}, 5385 (1999).

\bibitem{aac10} A. Acin, R. Augusiak, D. Cavalcanti, C. Hadley, J. K. Korbicz,
M. Lewenstein, L. Masanes, and M. Piani, Phys. Rev. Lett. {\bf
104}, 140404 (2010).

\bibitem{dxy10} R. Duan, Y. Xin, and M. Ying, Phys. Rev. A \textbf{81},
032329 (2010).

\bibitem{UPBother} D. P. DiVincenzo, T. Mor, P. W. Shor, J. A. Smolin, and B. M. Terhal, Comm. Math. Phys. \textbf{238}, 379
(2003).

\bibitem{note1} We mention a curious, maybe accidental, fact. If one could replace
the factors $\frac{1}{2}$ and $\frac{1}{2\sqrt{2}}$ with $1$ in
(\ref{explrho}), the resulting correlations would exhibit remarkable
non-local features: see Eq.(9) (up to bit flips) in V. Scarani, AIP Conference
Proceedings \textbf{844}, 309 (2006) and arXiv:quant-ph/0603017v2.
Unfortunately, the replacement would not define a valid quantum
state. The non-locality of bound entangled states under single-copy
measurements (Peres' conjecture) remains unsolved at the moment of
writing.


\bibitem{hub09} R. H\"ubener, M. Kleinmann, T.-C. Wei, C. Gonz\'alez-Guill\'en, and O. G\"uhne, Phys. Rev. A \textbf{80}, 032324
(2009).

\bibitem{markham10} D. Markham, arXiv: 1001.0343 (2010).

\bibitem{footnote_symmetry}
Note already that, as it was the case for the states $\ket{\varphi_i}$, and as it will be the case for all the bases we shall construct in this paper, the states $\ket{\phi_i}$ have the following symmetry: $\ket{\phi_0}$ is symmetric, while $\ket{\phi_1}$,$\ket{\phi_2}$ and $\ket{\phi_3}$ are obtained from one another by cyclic permutations $A \to B \to C \to A$.

\bibitem{ys06} The state had only been charecterized so far with another entanglement measure, proposed by
C. Yu and H. Song,  Phys. Rev. A {\bf 73}, 022325 (2006).


\end{thebibliography}

\end{document}